\documentclass[%
 reprint,
 amsmath,amssymb,
 aps,
]{revtex4-2}

\usepackage{caption}
\usepackage{subcaption}
\usepackage{amsmath}
\usepackage{url}
\usepackage{graphicx}% Include figure files
\usepackage{dcolumn}% Align table columns on decimal point
\usepackage{bm}% bold math
\begin{document}

\preprint{APS/123-QED}

\title{Non-selective Evaporation of Ethanol-Water Binary Mixture within Heated Capillary}

\author{Jialing Yu\textsuperscript{1}}
\author{Zhenhai Pan\textsuperscript{2}}
\email{E-mail address: panzhh\_sit@163.com (Zhenhai Pan).}

\affiliation{\textsuperscript{1}School of Mechanical Engineering, Shanghai Jiao Tong University, Shanghai 200240, China
\\\textsuperscript{2}school of Mechanical Engineering, Shanghai Institute of Technology, Shanghai 201418, China}

\date{\today}

\begin{abstract}

In the study, the evaporation of ethanol-water binary mixture within heated capillary is experimentally and numerically investigated. It was found that the ratio of the evaporation rates of ethanol and water equals the ratio of their initial concentrations in the mixture. This observation contradicts the commonly accepted view of selective evaporation, where the ratio of ethanol-to-water evaporation rates is expected to be  considerably higher than the concentration ratio owing to the higher volatility of ethanol. We term this novel phenomenon as non-selective evaporation. Subsequently, through numerical study, it was discovered that the changes in component concentration induced by the ethanol preferential evaporation appear solely in the limited area, referred to as the diffusion layer, near the meniscus. When the diffusion layer is fully developed, the evaporation process will transition from selective evaporation stage to non-selective evaporation stage. Due to the short duration of the selective evaporation stage in current study, the evaporation process exhibits distinct characteristics of non-selective. Moreover, by considering the coupled effects of convection and diffusion, an analytical model was proposed, and the criteria related with \textit{Pe} number were established to determine whether the evaporation process of binary mixture exhibits selective or non-selective characteristics. 

\end{abstract}

\maketitle

The evaporation of binary liquids is a ubiquitous phenomenon, occurring in both natural environments and industrial processes. Simultaneously, the evaporation of binary liquids inherently involves intricate heat and mass transfer mechanisms and exhibits complex evaporative behaviors, which has led to a growing academic interest in the evaporation of binary mixture over the past years 
 \cite{RN1,RN2,RN3,RN4,RN5}.
\par Due to differences in volatility among components, there exists an inevitably distinction between the ratio of evaporation rate of two components and the ratio of their concentration in the binary mixture, a phenomenon commonly called as “selective evaporation” 
 \cite{RN6,RN7,RN8}. The selective evaporation will result in a temporal variation in the composition, with a reduction in the concentration of the more volatile component \cite{RN9,RN10,RN11}. It is worth noting that the rate of decrease in the concentration of more volatile components is not spatially uniform. When the spatial concentration differences induced by selective evaporation cannot be offset by convective and diffusive mixing effects, there will present a distinct chemical separation phenomenon within the binary liquid, as observed in the experiment \cite{RN12,RN13,RN14,RN15}.
\par Within the binary liquid, the temporal-spatial variation of component concentration will directly affect the internal convection and evaporation rate. Therefore, it is generally considered that the evaporation process of binary mixture is a typical unsteady process 
 \cite{RN16,RN17,RN18}. 
In prior investigations, researchers predominantly focused on sessile binary droplet as the study object and yielded numerous intriguing findings  \cite{RN19,RN20,RN21}. Besides, the evaporation of binary mixture within constrained space, such as capillary, equally holds significant industrial relevance, yet remains an inadequately comprehended phenomenon \cite{RN22,RN23}.

\par In this study, experimental and numerical investigations were conducted for ethanol-water mixture evaporation within a heated capillary.  The evaporation process of binary mixture is described in detail by introducing the concept of the diffusion layer. And the effects of ethanol preferential evaporation, convective mass transfer, and diffusion mass transfer on the spatio-temporal distribution of components concentration were discussed.
\par
\begin{figure}[h]

\begin{center}
\includegraphics[width=0.46\textwidth]{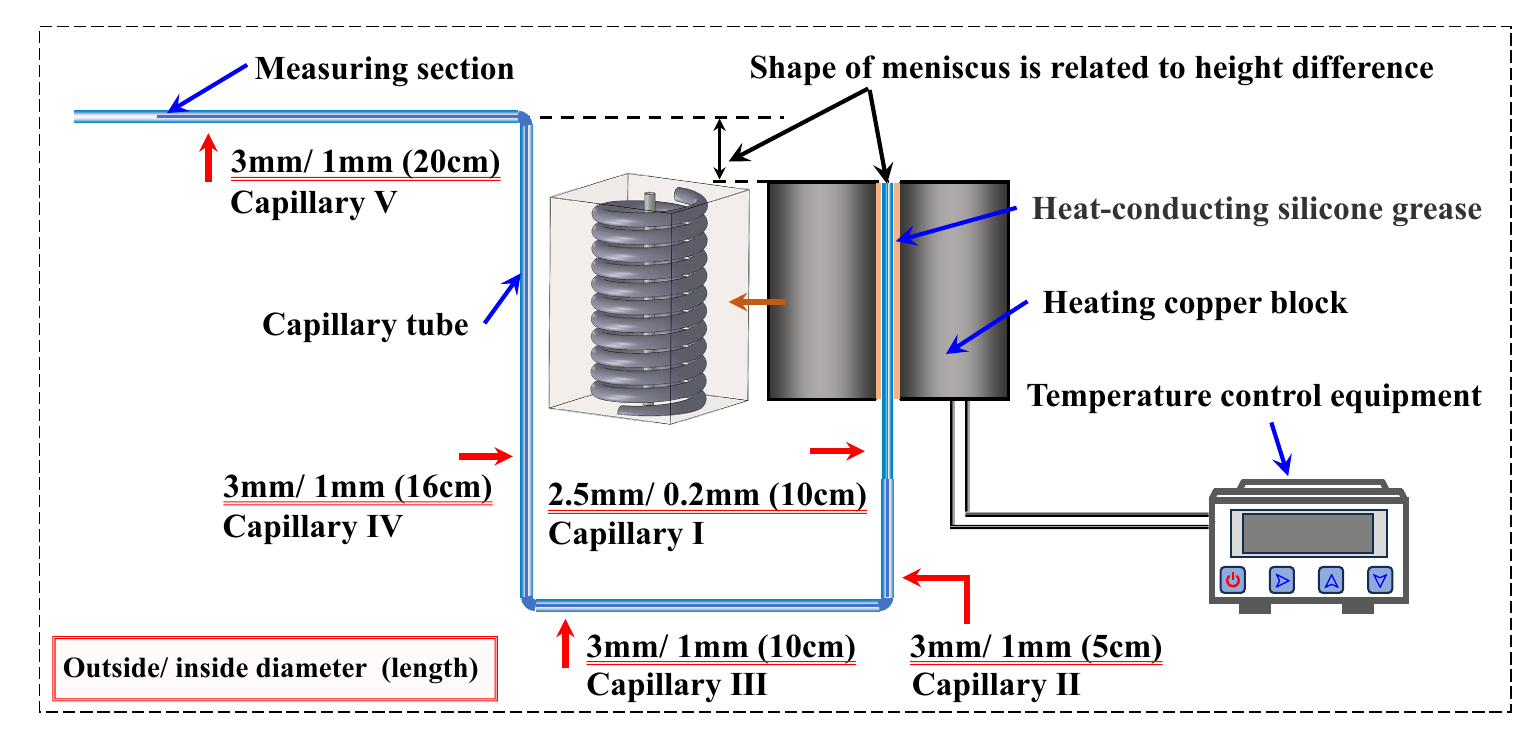}

\end{center}
\caption{\label{fig1} Schematic of the experimental setup.}

\end{figure}

 In the experiment setup, a binary mixture with 90\% mass concentration of ethanol and 10\% mass concentration of water evaporates in a customized capillary consisting of five straight capillary tubes, as illustrated in Fig.~\ref{fig1}. Capillary \uppercase\expandafter{\romannumeral1} is heated through a copper block maintained at a temperature of 75\textsuperscript{o}C. The copper block incorporates a central cylindrical heating hole specifically designed for accommodating the capillary.

\par In the initial stage, the capillaries \uppercase\expandafter{\romannumeral1}, \uppercase\expandafter{\romannumeral2}, \uppercase\expandafter{\romannumeral3}, and \uppercase\expandafter{\romannumeral4} are filled with the ethanol-water mixture. However, within the capillary \uppercase\expandafter{\romannumeral5}, there is a noticeable distinction between the meniscus and the end of capillary. Furthermore, the temperature of capillary \uppercase\expandafter{\romannumeral1} is noticeably higher than that of capillary \uppercase\expandafter{\romannumeral5}. Therefore, it is reasonable to neglect the evaporation of the binary liquid at the meniscus within the capillary \uppercase\expandafter{\romannumeral5}, assuming that evaporation solely occurs at the meniscus in the capillary \uppercase\expandafter{\romannumeral1} \cite{RN24}.  
\par
\begin{figure}[h]

\includegraphics[width=0.47\textwidth]{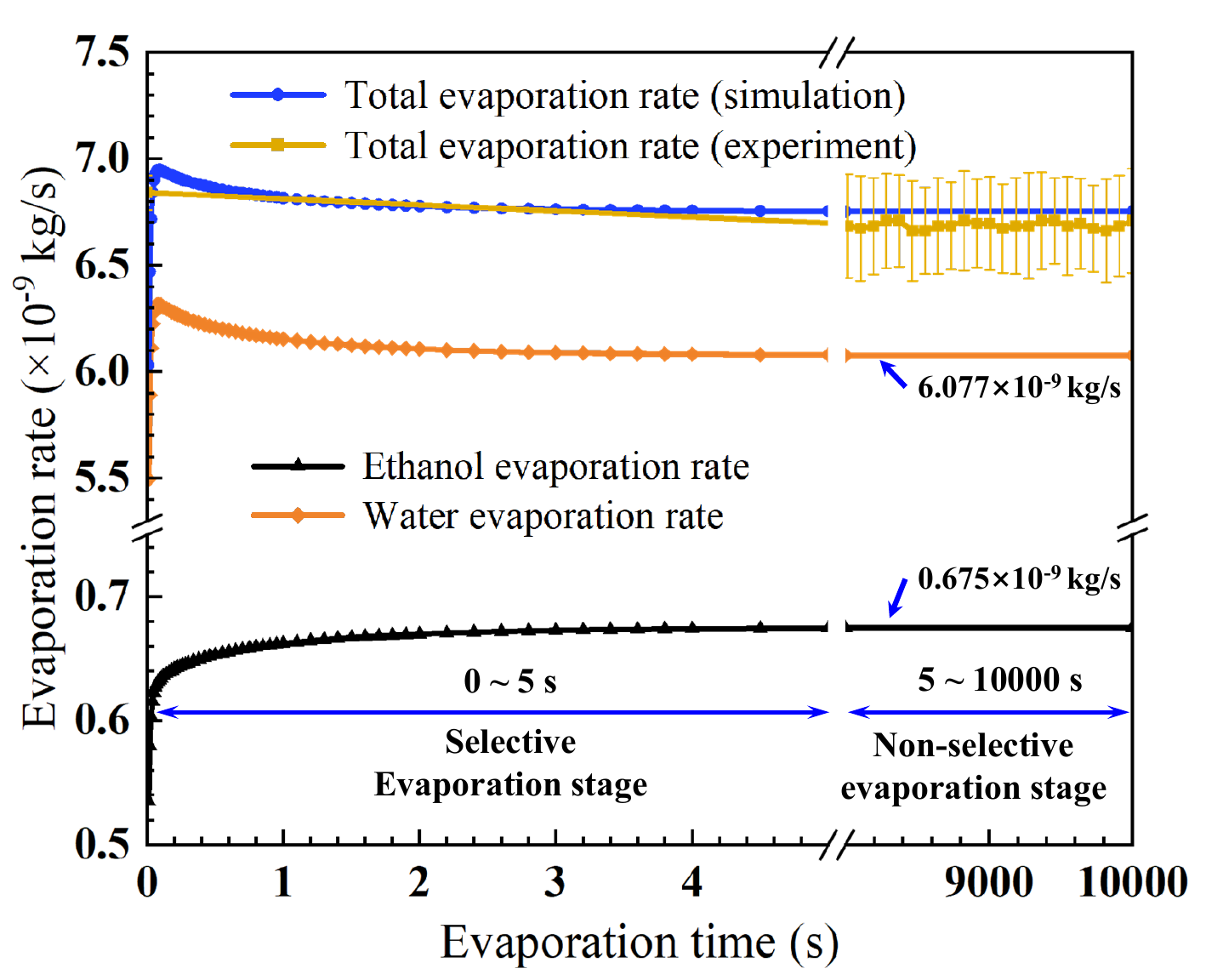}

\caption{\label{fig2} Temporal variation of ethanol, water, and total evaporation rate of ethanol-water mixture in heated capillary.
}

\end{figure}

 Under the effect of capillary force, evaporation loss of binary mixture at the meniscus in the capillary \uppercase\expandafter{\romannumeral1} will be compensated from the capillary \uppercase\expandafter{\romannumeral5}. The meniscus in capillary \uppercase\expandafter{\romannumeral1} is pinned at the upper end and the meniscus within the capillary \uppercase\expandafter{\romannumeral5} continuously migrates in the direction away from the end during the evaporation process. By quantifying the migration rate of the meniscus within the capillary \uppercase\expandafter{\romannumeral5}, it is possible to deduce the evaporation rate occurring at the meniscus within the capillary \uppercase\expandafter{\romannumeral1}. The result indicated that the total evaporation rate of mixture remains nearly constant throughout the evaporation process, as dipicted in Fig.~\ref{fig2}. This phenomenon differs from the widely accepted perspective, wherein the total evaporation rate of binary mixture is generally considered to gradually decrease \cite{RN25,RN26}. This expectation arises from the fact that the more volatile component tends to evaporate preferentially, causing a decrease in its concentration and subsequently slowing down the total evaporation rate.
\par Moreover, Karl Fischer moisture titrator (KFT) is employed to measure the average water concentration of the binary mixture across multiple experiments. All experimental conditions remain consistent, the only variation lay in the duration of the evaporation process. After completing any single experiment, the remaining mixture within the capillary was transferred to small vials and immediately sealed for subsequent KFT analysis. The water concentration exhibits minimal variation over multiple experiments with different evaporation durations. This phenomenon also contradicts the commonly accepted viewpoint.
\par Compared to the selective evaporation which is commonly observed in binary mixture  \cite{RN17,RN27}, the characteristic of observed novel phenomenon is that the ratio of evaporation rate of the two components is consistent with the ratio of their concentrations in the mixture, independent of volatility difference between two components. We term this evaporation phenomenon as non-selective evaporation because the volatile component does not preferentially evaporate in this process. To the best of our knowledge, the phenomenon of non-selective evaporation has not been previously observed or investigated. 
\par To reveal the underlying physics mechanism of the non-selective evaporation, and analyze the temporal and spatial variations in the component concentration, a numerical model is developed.
\par  To reduce the computational cost, our model focuses exclusively on capillary \uppercase\expandafter{\romannumeral1} and its adjacent air environment, incorporating an assumption of axisymmetry. The ethanol-water mixture evaporates at the meniscus fixed on the upper end of capillary, while being replenished from the opposite end. The meniscus is assumed to be spherical, with a constant contact angle of 60\textsuperscript{o}. The wall temperature of the capillary is fixed at 75\textsuperscript{o}C. See Supplemental Material for the detailed specifics of the numerical model.

\par Initially, the capillary is filled with water-ethanol mixture (90\% ethanol mass fraction) at room temperature (25\textsuperscript{o}C).
The temporal variations in the individual evaporation rates of ethanol and water, as well as the total evaporation rate, are depicted in Fig.~\ref{fig2}. The simulation results exhibit excellent consistency with the experimental data.

\begin{figure}[h]
\begin{center}
\includegraphics[width=0.47\textwidth]{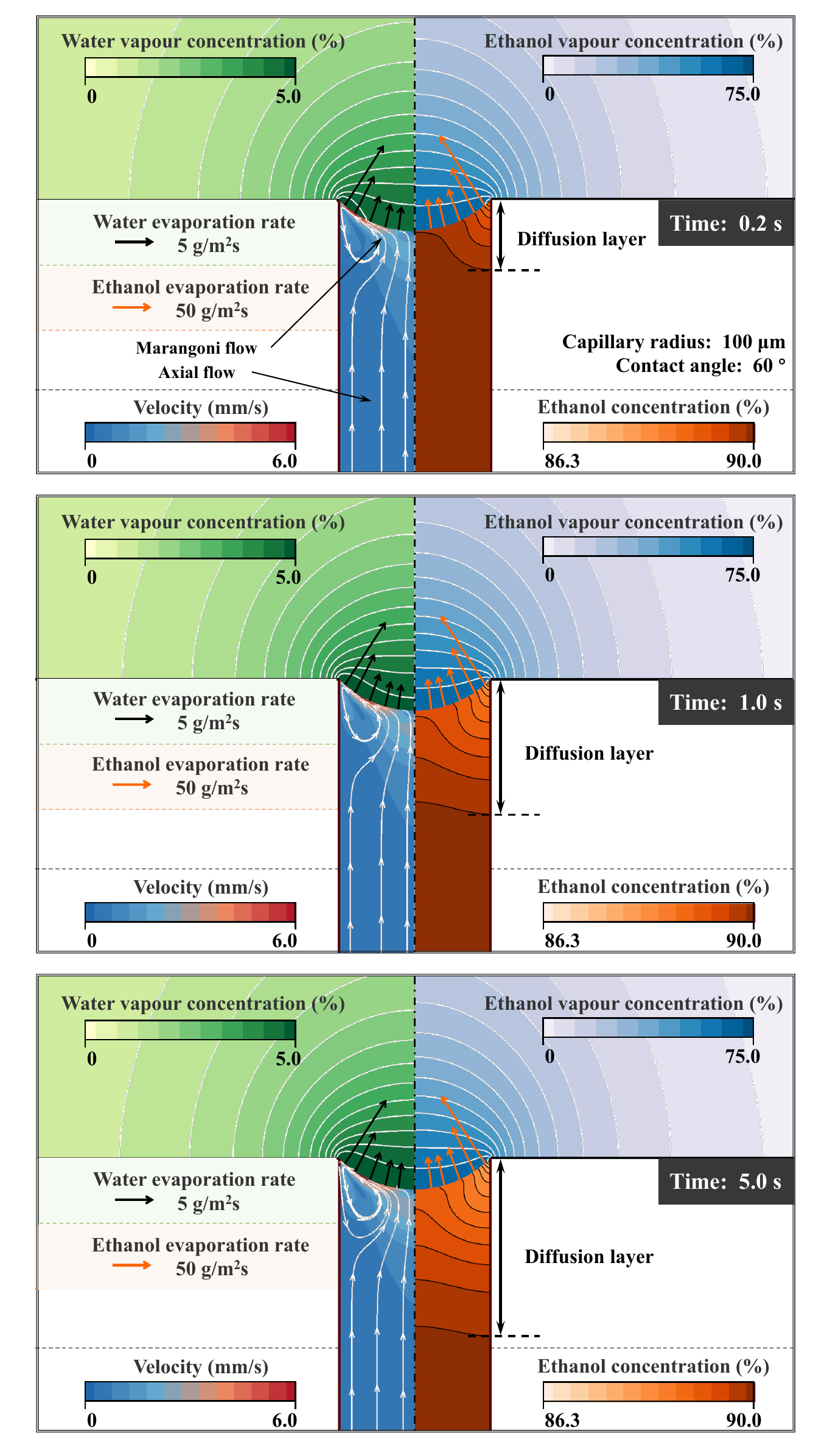}

\end{center}
\caption{\label{fig3} The snapshots of evaporation process of water–ethanol mixture in capillary, initially containing a 90\% mass fraction of ethanol, at three different times t = 0.2s , t = 1.0s  and t = 5.0s . In the liquid phase,  the mass fraction of ethanol (right) and velocity (left) is sketched. In the gas phase, vapour concentration of ethanol (right) and water (left) is depicted. And the evaporation rate of ethanol (right) and water (left) is pointed out by the arrows at the meniscus.}

\end{figure}

\par The variations of the temperature, flow fields, vapor distributions in the gas, and component concentrations of the ethanol are shown in Fig.~\ref{fig3}. In the beginning, the ethanol concentration at the meniscus gradually decreases due to its higher volatility. The changes in the mixture composition gradually diffuse from the meniscus towards the interior of the capillary. Meanwhile, the concentrated water at the meniscus suppresses the evaporation of ethanol but enhances the evaporation of water. And the total evaporation rate gradually decreases. However, this variation continues for only ~5 seconds, after which the evaporation process becomes steady.

\par It worth to notice that, during the steady state of evaporation, the ratio of the evaporation rates of ethanol and water precisely matches the initial mass fraction ratio of the two components in the binary mixture (6.077×10\textsuperscript{-9}kg/s / 0.675×10\textsuperscript{-9}kg/s $\approx $ 9/1).

\par The simulation results show that the evaporation process of the binary mixture in heated capillary can be divided into two stages comprising an initial brief stage of selective evaporation followed by subsequent non-selective evaporation stage. The boundary between the two evaporation stages is the moment when the evaporation process of the binary mixture reaches a steady state. During the stage of selective evaporation, the total evaporation rate slows down, and the ethanol concentration decreases which is consistent with generally accepted views. However, the selective evaporation stage only exists for a short time (approximately 5 seconds) compared to the total evaporation time (approximately 10000 seconds). Therefore, the selective evaporation stage and its effects were not observed in the experiment.

\par Due to the evaporation process reaches a steady state in a short time, and the duration of selective evaporation is significantly shorter compared to non-selective evaporation. The evaporation process of ethanol-water mixture in heated capillary exhibits the obvious characteristics of non-selective evaporation. It can be inferred that the key criterion for determining whether the evaporation process of a binary mixture is selective or non-selective lies in its potential to attain steady state. If steady state is attainable, one also needs to consider the ratio between the time for the evaporation process to stabilize and the total evaporation time. Only when the time for the evaporation process to stabilize is significantly shorter than the total evaporation time, the evaporation process of a binary mixture exhibits the characteristics of non-selective evaporation. 

\par During the evaporation process, the ethanol preferential evaporation changes the mixture composition at the meniscus and transfers such variation to the binary mixture inside the capillary by diffusion. Here, we refer to the area where the liquid composition changes as the diffusion layer, and the axial distance of the diffusion layer as the diffusion length.

\par At the same time, there is an axial convection from the capillary interior toward the meniscus existing in the liquid phase, which serves to replenish the liquid lost due to evaporation. The axial convection velocity is proportional to the total evaporation rate of the binary mixture. The direction of axial convection is opposite to the extension direction of the diffusion layer, tending to impede the extension of the diffusion layer from meniscus into the capillary interior. 

\par Under the coupling effect of diffusion and axial convection, the evolution of the diffusion layer during the evaporation process can be analyzed. When the ratio of the diffusion rate to the total evaporation rate is small, the influence of the preferential evaporation on the liquid composition will be limited to a specific range at a certain length from the meniscus and the evaporation process can reach a steady state. This certain length is called as ultimate diffusion length.

\par The ultimate diffusion length can be estimated by a simplified one-dimensional steady analytical model, which is described by the following governing equation:
\[{{D}^{l}}\frac{{{\partial }^{2}}{{w}_{e}}}{\partial {{z}^{2}}}-u\frac{\partial {{w}_{e}}}{\partial z}=0\]
\par At the replenishment end, the ethanol concentration is equal to the initial concentration ${{w}_{e}}={{w}_{e,i}}$, at $z=-L$, where $L$ denotes the length of capillary. At the capillary evaporation end, ${{w}_{e}}={{w}_{e,f}}$, at $z=0$, and ${{w}_{e,f}}$ can be calculated from the fact that the ratio of the evaporation rates of ethanol and water is precisely equal to the ratio of the initial concentrations in the mixture during steady state.

\begin{figure}[h]
  \centering
  \begin{subfigure}{0.48\textwidth}
    \includegraphics[width=0.85\textwidth]{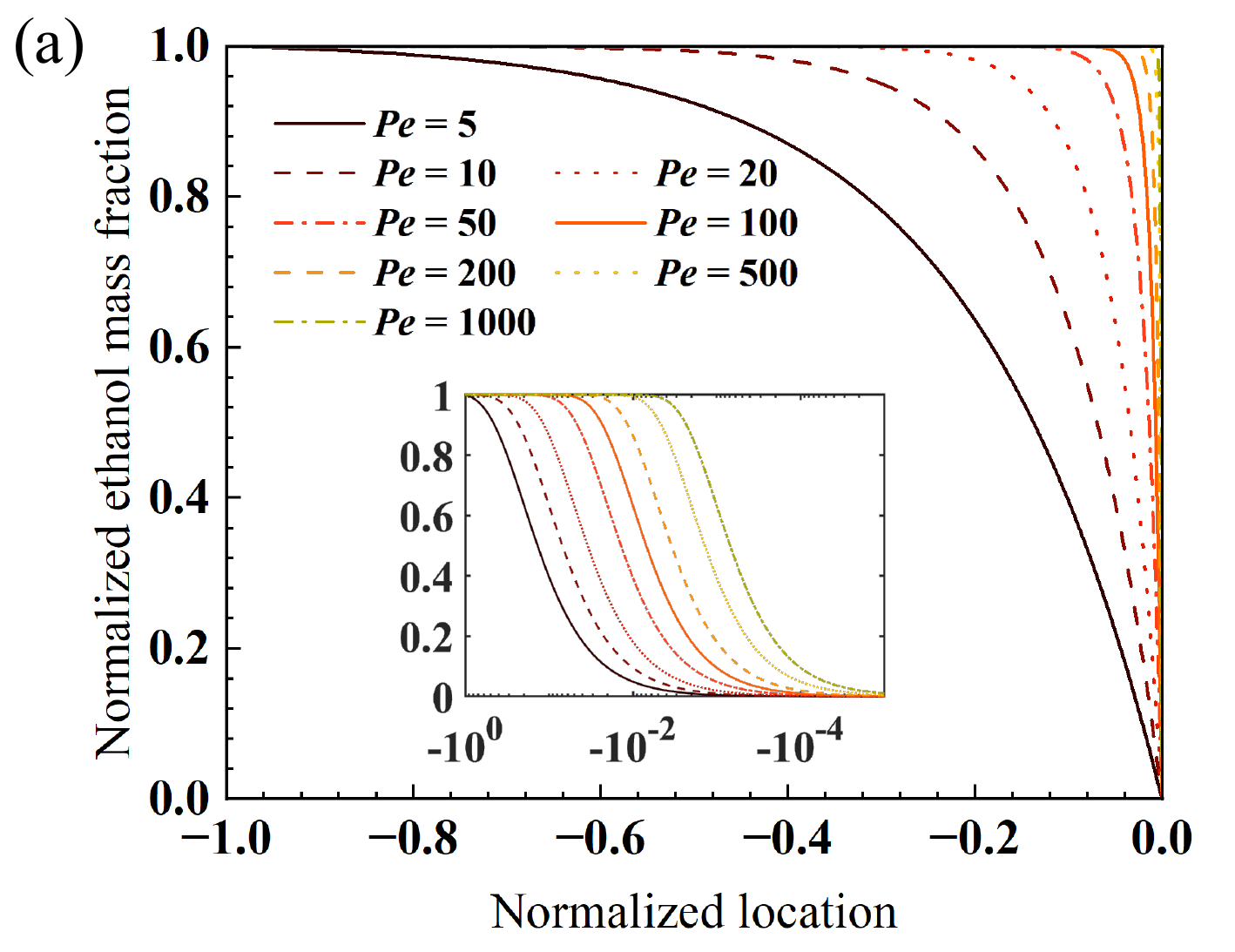}
    \label{fig4a}
  \end{subfigure}
  \vfill
 
  \begin{subfigure}{0.48\textwidth}
    \includegraphics[width=0.85\textwidth]{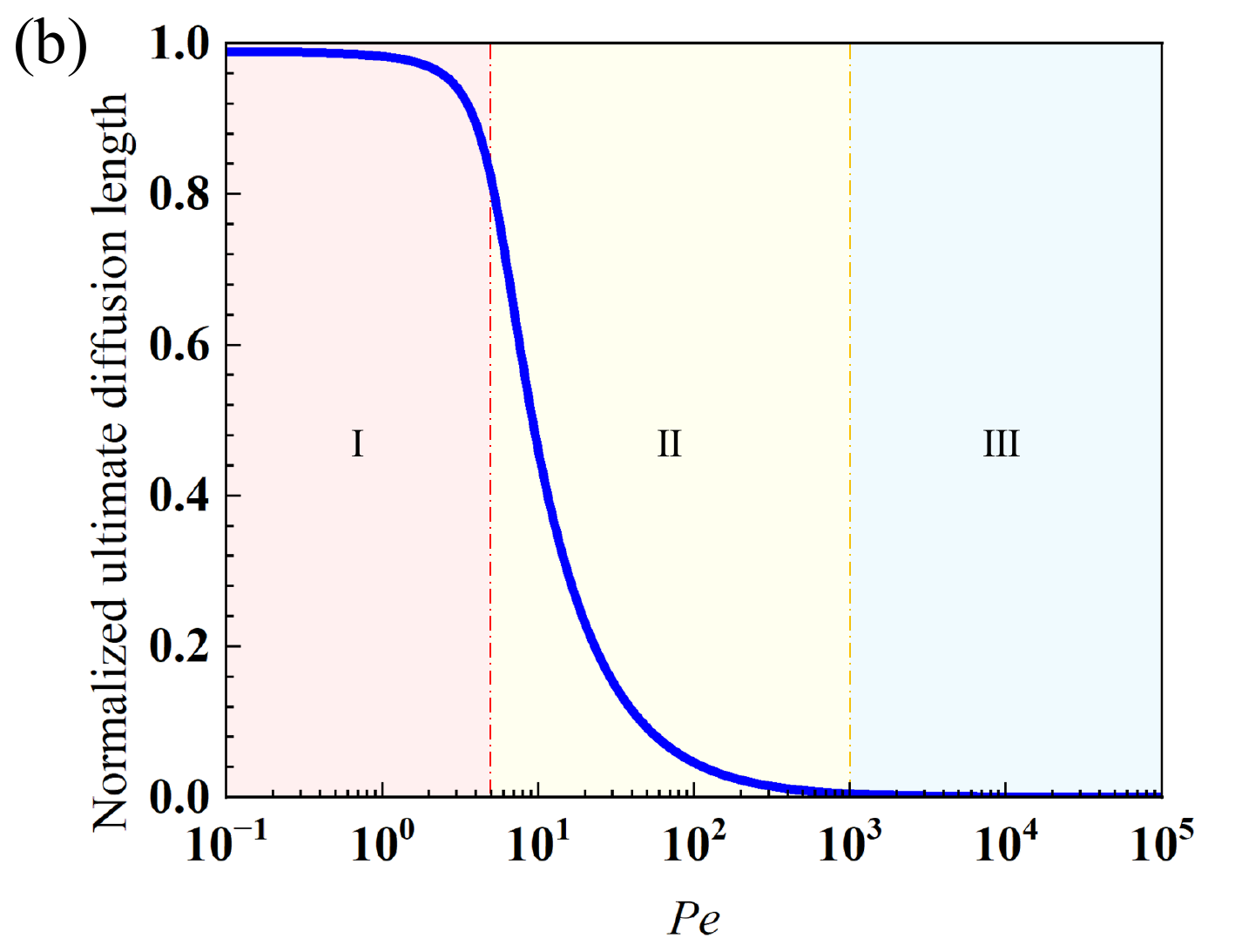}
    \label{fig4b}
  \end{subfigure}
 
  \caption{(a) Distribution of normalized ethanol mass fraction under different \textit{Pe} number and (b) normalized ultimate diffusion length as a function of \textit{Pe} number }
  
  \label{fig4}
\end{figure}

\par With normalized variables of ethanol concentration ${{\bar{w}}_{e}}=\left( {{w}_{e}}-{{w}_{e,f}} \right)/\left( {{w}_{e,i}}-{{w}_{e,f}} \right)\ $ and length $\bar{z}=z/L\ $, the normalized governing equation and boundary conditions are given as follow:
\[\frac{{{\partial }^{2}}{{{\bar{w}}}_{e}}}{\partial {{{\bar{z}}}^{2}}}-Pe\frac{\partial {{{\bar{w}}}_{e}}}{\partial \bar{z}}=0\]
\par The boundary conditions are as follows: at $\bar{z}=-1$, the normalized ethanol concentration is fixed at ${{\bar{w}}_{e}}=1$, and at $\bar{z}=0$, ${{\bar{w}}_{e}}=0$.

\par Where the Peclet (\textit{Pe}) number is defined as $Pe={uL}/{{{D}^{l}}}\;$, denoting the ratio of axial velocity to diffusion velocity.

\par Solve the formula, the normalized ethanol concentration is expressed as:
\[{{\bar{w}}_{e}}=\frac{1}{{{e}^{-Pe}}-1}{{e}^{Pe\cdot \bar{z}}}+\frac{1}{1-{{e}^{-Pe}}}\]
\par Fig.~\ref{fig4}a illustrates the spatial distribution of ethanol concentration inside the capillary during steady state under different \textit{Pe} numbers. It can be seen that the larger the  \textit{Pe} number, the larger the area where the concentration of the mixture components in the capillary changes. We consider the area with normalized ethanol concentration of less than 0.99 as the diffusion layer. The normalized length of the ultimate diffusion layer is negative correlated to the \textit{Pe} number, which is dependent on the ratio of diffusion velocity to axial convection velocity, as depicted in Fig.~\ref{fig4}b.

\par The smaller the \textit{Pe} number, the shorter the ultimate diffusion length, and the less time is required to reach a steady state, and vice versa. For instance, when the \textit{Pe} number exceeds 1000  (region \uppercase\expandafter{\romannumeral3} in Fig.~\ref{fig4}b), it can be considered that the ultimate diffusion distance is negligible compared to the capillary length. Hence, the selective evaporation stage is also negligible. In this case, the evaporation process is markedly non-selective.

\par On the contrary, when the ratio of the diffusion rate to the total evaporation rate is large, it can be considered that \textit{Pe} number is less than 5 (region \uppercase\expandafter{\romannumeral1} in Fig.~\ref{fig4}b). The change of the liquid composition will diffuse into the entire binary mixture from meniscus. The concentration of ethanol (more volatile component) in the mixture and total evaporation rate will persistently decline, preventing the evaporation process from reaching a steady state. In this case, the evaporation process will exhibit the characteristics of selective evaporation. Limited by the boundary conditions at the replenishment end of the capillary, the normalized ultimate diffusion length will not be greater than 1. In fact, in this situation, the ultimate diffusion distance is greater than the characteristic length of the binary mixture, but it cannot be observed intuitively.  

\par Except for preferential ethanol evaporation and diffusion, the Marangoni convection at the meniscus also has a significant impact on the spatial distribution of component concentrations as depicted in Fig.~\ref{fig3}. Marangoni convection accelerates the mixing of binary mixture within the flow range, resulting in a relatively uniform spatial distribution of ethanol and water concentrations. It can be seen that in areas with a strong vortex flow, the spatial distribution of component concentration is mainly affected by convection rather than diffusion and axial convection. 

\par Based on the above analysis, the reasons for the evaporation process of ethanol-water mixture within the heated capillary reaching steady in a short time are as follows: 1) Apart from the Marangoni convection at the meniscus, constrained by the small capillary diameter, there is no other vortex flow inside the capillary. The component concentration of the majority binary mixture within the capillary is primarily dominated by axial convection and diffusion; 2) Under the temperature of 75\textsuperscript{o}C, the total evaporation rate of the binary mixture within the capillary is significantly higher than the mutual diffusion rate of ethanol and water \cite{RN28}. Resulting in a small ultimate diffusion length compared to the capillary length, thereby a short time to reach a steady state is required.

\par In summary, the non-selective evaporation is observed in the evaporation process of ethanol-water mixture in the heated capillary. And the reason for this novel phenomenon is explained detailedly. Our findings provide new insights for comprehending the evaporation process of binary mixtures. Moreover, it provides a method where all miscible mixtures can have the same composition ratio between the liquid phase and vapor generated during vaporization, not limited to azeotropic mixtures.

%\bibliography{main}% Produces the bibliography via BibTeX.


\end{document}